# Polygonal Spatiotemporal Optical Vortices Wavepackets with Prescribed Vortex Structure

Haihao Fan[1,†], Qian Cao[1,†,*], Andy Chong[3,4], and Qiwen Zhan[1,2,5,*]

[1] School of Optical-Electrical and Computer Engineering, University of Shanghai for Science and Technology, Shanghai 200093, China.

[2] Zhejiang Key Laboratory of 3D Micro/Nano Fabrication and Characterization, Department of Electronic and Information Engineering, School of Engineering, Westlake University, Hangzhou 310030, China

[3] Department of Physics, Pusan National University, Busan 46241, Republic of Korea

[4] Institute for Future Earth, Pusan National University, Busan 46241, Republic of Korea

[5] Westlake Institute for Optoelectronics, Fuyang District, Hangzhou, Zhejiang 311400, China

† These authors contribute equally to this work.

*cao.qian@usst.edu.cn, qwzhan@usst.edu.cn

**Abstract:** Optical vortices are a type of light beams that carry orbital angular momentum(OAM), enabling the generation and control of additional degrees of freedom. According to the orientation of orbital angular momentum, optical vortices can be classified into spatial optical vortex beam carrying longitudinal OAM (L-OAM) and spatiotemporal optical vortices carrying transverse OAM (T-OAM). As an emerging subset of optical vortices, polygonal optical vortices provide a unique platform for a wide range of frontier applications by introducing a new degree of freedom in the form of a customizable intensity structure. In the spatial domain, polygonal spatial optical vortex beam carrying L-OAM have already demonstrated great potential in optical manipulation and two-photon lithography. However, polygonal spatiotemporal optical vortex (STOV) wavepackets contain multiple sub–STOVs carrying T-OAM remains unrealized to date. In this work, we theoretically propose and experimentally demonstrate polygonal spatiotemporal optical vortices wavepackets embedded with prescribed vortex structures. Within the structure, a prescribed number of sub-STOVs carrying T-OAM is set along a designed polygonal spatiotemporal trajectory. Using the spatiotemporal holographic shaping approach, we generate polygonal perfect STOV wavepacket and use the combination of multiple polygonal perfect STOV wavepacket to form polygonal STOV wavepacket with the prescribed vortex structure. A full control over multiple key properties of the polygonal STOV wavepackets such as the geometry, number of phase singularities, and spatiotemporal distribution of sub-STOVs is also achieved. These new spatiotemporal wavepackets will facilitate applications such as novel optical communication, study of complicated quantum systems, and multiple-target particle manipulation.

## 1. Introduction

Structured light finds diverse applications owing to its degrees of freedom in both spatial and spatiotemporal structure[1-3]. As a type of structured light, optical vortices(OVs) is a "doughnut-shaped" beam with a helical phase wavefront and a central core of zero intensity. In the early 1990s, Allen *et al*. demonstrated that the longitudinal orbital angular momentum (OAM) of a photon is inherently carried by a spatial vortex beam, with the magnitude of the OAM being proportional to its topological charge(TC)[4,5]. Since then, optical vortex beams carrying longitudinal OAM, including Laguerre–Gaussian beams, Bessel–Gaussian beams, perfect optical vortices beam and others, have been extensively demonstrated and applied in a wide range of applications, such as optical communications[6,7], optical tweezers[8,9] and quantum information [10,11]. In these applications, the helical phase and intensity distribution of the optical vortex play a crucial role. Therefore, increasing demand for practical applications is forcing more profound research into OVs: from the generation and measurement to the shaping and multiple vortices manipulation of OVs[2,12-14].

To date, several methods have been proposed to generate polygonal optical vortex beam, such as using the high-order cross-phase[15,16], optical pen[17], astigmatic transformation[18], free lens modulation[19], and mode-locked quasi-frequencydegenerate laser[20]. These methods are used to generate polygonal optical vortex beam with arbitrary polygonal structures, and the switching between different polygonal structures has also been demonstrated. Unlike traditional circular optical vortex beams, polygonal optical vortex beam refers to vortex beams carrying OAM with a closed polygonal pattern in intensity distribution. These polygonal optical vortex beam with arbitrary polygonal profiles provide additional possibilities in advanced applications, such as optical manipulation[21,22], single-shot lithography[23], and optomechanical assemblies[24,25]. In contrast to conventional vortex beams with single vortices, to meet the requirements of complex applications, recent advances highlight the optical vortex lattices with multiple vortices and various geometric shapes[26-30]. Typically, these optical vortex lattices with various geometric shapes are generated via the superposition of polygonal optical vortices beams[2]. These optical vortex lattices can considerably enhance the optical communication capacity and speed[31,32], as well as enable the complex manipulation of multiple particles[29,33,34]. However, the generated optical vortex lattices are still limited to spatial vortical patterns, implying that the multiple vortices present within them carry longitudinal orbital angular momentum.

Recently, there has been intense interest in exploring the spatiotemporal forms of optical vortices. The study of spatiotemporal optical vortices (STOV) wavepackets has made great progress with the help of temporal shaping and spatial shaping techniques[35-40] by controlling the two-dimensional spatial light modulator that is placed in a pulse shaper to generate the STOV wavepacket that can carry a transverse orbital angular momentum (T-OAM)[41,42]. As a new degree of freedom for optical manipulation, many research groups have started to study novel spatiotemporal optical wavepackets such as spatiotemporal Bessel[43,44], crystal[45], Laguerre and Hermite Gaussian[46,47] and circular airy wavepackets[48]. Recently, our research group have proposed and demonstrated a perfect STOV wavepacket whose wavepacket size is independent of the topological charge, addressing the limitation of traditional STOV wavepackets that its spatiotemporal spread increases with an increasing topological charge, and we constructed a circular STOV wave packet with multiple vortices distributed along a circular arrangement[49]. Although significant progress has been made in the research on STOV wavepackets, the currently generated STOV wavepackets, whether with a single STOV or multiple sub-STOVs, typically exhibit a single circular symmetric structure in their intensity distribution, and lack the dynamic tunability in the spatiotemporal structure, resulting in the failure to fully utilize their potential in spatiotemporal intensity information. These factors can limit the use of this wavepacket in complicated applications such as multiple-target particle tweezering and imaging systems.

In this context, we propose a method to construct polygonal STOV wavepackets embedded with a controllable number of sub-STOVs, each of which carries T-OAM. The scheme is based on the combination of two polygonal perfect STOV wavepackets, whose radial size and shape can be freely controlled, thereby enabling precise arrangement of the sub-STOVs along arbitrary prescribed polygonal path. Employing spatiotemporal holographic shaping technology, we experimentally demonstrated flexible control over the geometric shape of the polygonal spatiotemporal vortex wavepacket, the number of phase singularities, and the spatiotemporal distribution of the sub-STOVs. This new polygonal STOV wavepackets together with the customizable sub-vortices will become an interesting tool in applications such as novel optical communication, quantum physical studies, and particle manipulations.

## 2. Theoretical analysis and numerical simulations

Polygonal STOV wavepackets with a prescribed vortex structure can be constructed by combining two individual polygonal STOV wavepackets. This composition can be understood as the spatiotemporal counterpart of superposing multiple beams for generating an optical vortex lattice beam[37,38]. Here, we first investigate the creation of a polygonal STOV wavepacket. Its generation in the spatiotemporal domain ($X-T$ plane) can be achieved by introducing an azimuthal phase modulation in the spatial-spectral Fourier domain ($k_x - \omega$ plane). The azimuthal phase modulation correlates the vortex phase with the azimuthal angle position[50] so that the contours of the vortex phase can vary nonlinearly. The resulting nonlinearly varying azimuthal phase can be written as

$$\eta(\varphi) = l \cdot \varphi + \alpha \cdot \cos(\beta \cdot \varphi + \Psi_0), \tag{1}$$

where $(\rho, \varphi)$ are the corresponding polar coordinates in the spatial-spectral Fourier plane, $\rho = \sqrt{k_x^2 + \gamma^2 \omega^2}$, and $\varphi = \tan^{-1}(\gamma \omega / k_x)$. $\gamma$ is the scaling factor with a unit of ps/mm for controlling the aspect ratio for the $k_x - \omega$ plane. $\alpha$ is the nonlinear coefficient for controlling the smoothness of the polygon shape, and $\beta$ is an integer for controlling the number of sides of the polygon. orientation factor $\Psi_0$ controls the starting phase of the polygonal STOV wavepackets. The radius of the polygonal STOV wavepacket is related to the Bessel-Gaussian mode in the spatial-spectral Fourier plane. The Bessel-Gaussian mode can $\psi(\rho, \varphi)$ can be expressed by[49]

$$\psi(\rho, \varphi) = \exp\left(-\frac{\rho^2}{w_0^2}\right) \cdot J_l(k_r \rho) \cdot e^{-i\eta(\varphi)}, \tag{2}$$

where $w_0$ is the waist radius of the Bessel-Gaussian beam in the $k_x - \omega$ plane. $J_l$ is the first kind Bessel function with an order of $l$. $k_r$ is the radial wavevector of the Bessel-Gaussian beam. The size of the Bessel-Gaussian beam decreases as $k_r$ increases.

The phase variation of azimuthal angle has a boundary condition as $\eta(0) = \eta(2\pi)$ for constraining the value of $\beta$. The two-dimensional spatiotemporal Fourier transform (FT) of can transform a Bessel-Gaussian beam into a perfect vortex field, whose field can be expressed as

$$\begin{aligned}
\psi(r, \theta) &= \mathrm{FT}[\psi(\rho, \varphi)] \\
&= \mathrm{FT}\left[\exp\left(-\frac{\rho^2}{w_0^2}\right) \cdot J_l(k_r \rho) \cdot e^{-i\eta(\varphi)}\right] \\
&= \frac{1}{2\pi} \int_0^{2\pi} \int_0^\infty \exp\left(-\frac{\rho^2}{w_0^2}\right) \cdot J_l(k_r \rho) \cdot e^{-i\eta(\varphi)} \cdot e^{i\rho r \cdot \cos(\varphi - \theta)} \rho \, d\rho \, d\varphi \\
&\approx \frac{i^{l-1}}{k_r} \exp\left(-\frac{(r - r_0)^2}{w^2}\right) e^{-i\eta(\theta)},
\end{aligned} \tag{3}$$

where $r = \sqrt{X^2 + \alpha^2 T^2}$, $\theta = \tan^{-1}(X/\alpha T)$. $\alpha = \gamma^{-1}$ is the scaling factor in the spatiotemporal domain. The polygonal STOV wavepacket with a prescribed sub-STOVs structure can be generated by the superposition of two concentric polygonal perfect STOV wavepackets,

$$\psi_{\text{total}}(r, \theta) = \psi_1(r, \theta) \exp(i\phi_1) + \psi_2(r, \theta) \exp(i\phi_2), \tag{4}$$

where $\psi_1$ and $\psi_2$ are perfect STOV wavepackets with topological charge $l_1$ and $l_2$, respectively. $\Delta \phi = \phi_2 - \phi_1$ is the phase difference between these two perfect STOV wavepackets.

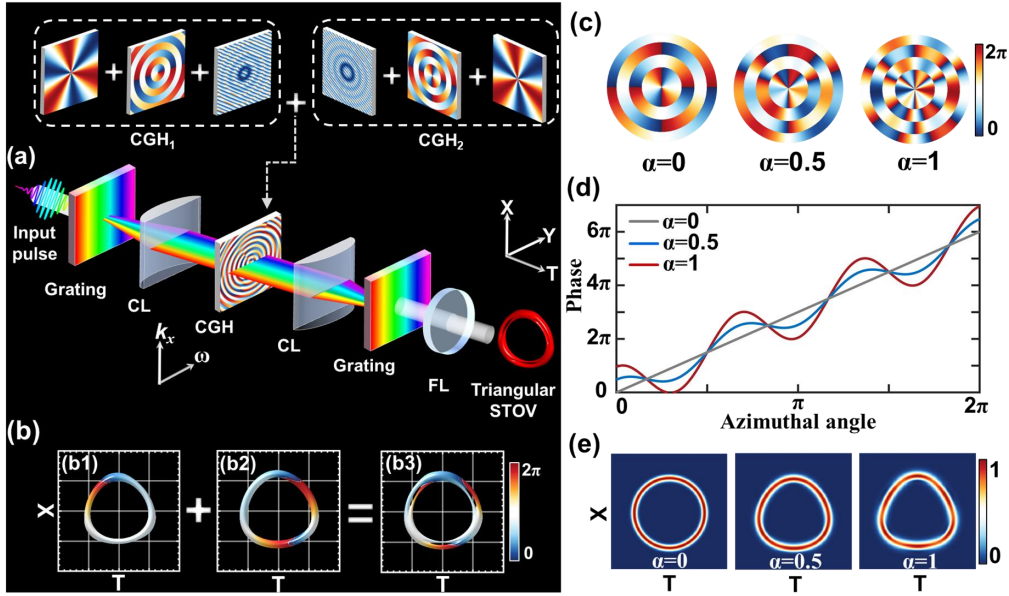

**Fig. 1. Concept of synthesizing a polygonal STOV wavepacket with a sub-STOVs structure.** (a) The conceptual scheme for synthesizing a polygonal STOV wavepacket using the spatiotemporal holographic pulse shaping technique. The inset plot shows that two computer-generated holograms (CGH) are encoded on the SLM for generating the polygonal STOV wavepacket with a triangular shape embedded with three sub-STOVs. (b) Spatiotemporal decomposition of the triangular STOV wavepacket. (c) Fourier domain phase for generating different triangular STOV. The nonlinear coefficient for the azimuthal phase $\alpha$ is set at 0, 0.5, and 1 for generating triangular STOVs with different smoothness. The transverse topological charge $l$ is set at +2; the number of sides $\beta$ of polygon is set at +3; (d) The azimuthal phase profile at different values of $\alpha$. (e) The resulting spatiotemporal intensity profile of the triangular perfect STOVs. CL: cylindrical lens.

The conceptual scheme for the synthesizing polygonal STOV wavepacket with a prescribed sub-STOVs structure is presented in Fig. 1(a). The core component of our synthesizer is to modulate the incident wavepacket by using the spatiotemporal computer-generated hologram (CGH), and then to generate the polygonal STOV wavepacket through the spatiotemporal Fourier transform[41]. Figure 1(a) shows the instance where the triangular STOV wavepacket is generated. To achieve this, two triangular perfect STOV wavepackets, as shown in Fig. 1(b), with different initial wavepacket sizes and different topological charges of $l_1 = +1$ and $l_2 = -2$ are generated. When the two triangular perfect STOV wavepackets are perfectly superposed with each other, the superposition of the wavepackets produces a triangular distribution of intensity nulls (dark dots) within the overlapped region. These dark dots are sub-STOVs within the wavepacket. The number of sub-STOVs can be calculated by $N = |l_2 - l_1|$. Here in Figure 1(b3), the number of sub-STOVs is $N = |l_2 - l_1| = |-2 - 1| = 3$. The phase profile around these sub-STOVs proves the existence of the local optical vortices in the dark spot.

To produce a polygonal perfect STOV wavepacket with different shapes, we change the phase of Bessel-Gaussian vortices in the spatial-spectral field by using the azimuth phase with a non-uniform phase gradient. The phase profile of Bessel-Gaussian vortices with $l = +2$, $\beta = 3$ and different nonlinear coefficient $\alpha$ is shown in Fig. 1(c). Compared with the Bessel-Gaussian vortices with $\alpha = 0$, the Bessel-Gaussian vortices with $\alpha = 0.5$ and 1 has a nonlinear variation in the phase profile [see Fig. 1(d)]. By loading the phase in Fig. 1(c) into SLM as a computer-generated hologram (CGH), different triangular perfect STOV light fields can be obtained. The results are shown in Fig. 1(e). It can be seen that as the nonlinear coefficient $\alpha$ increases, the intensity distribution on the vortex ring will gradually change into a triangular profile distribution.

## 3. Methods and results

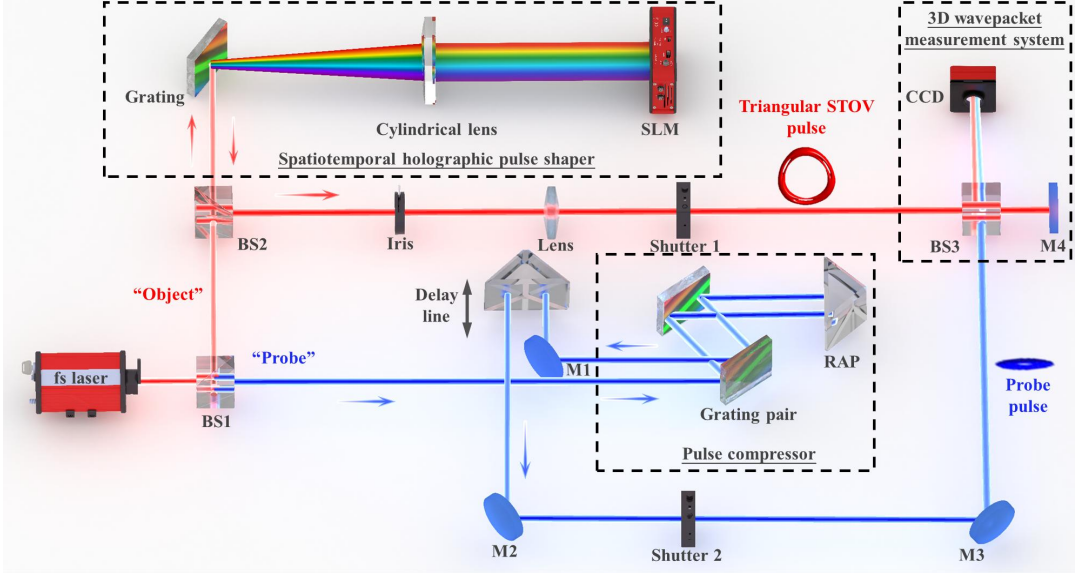

**Fig. 2. Experimental setup for synthesizing and characterizing polygonal STOV wavepackets.** The setup includes three sections: a spatiotemporal holographic pulse shaper, a pulse compressor using a grating pair, and a time delay line system for measuring the 3D intensity and phase profile of the generated polygonal STOV wavepackets. M: mirror; BS: beam splitter. RAP: right-angle prism.

Figure 2 shows the experimental setup for generating and characterizing the polygonal STOV wavepackets and the polygonal STOV wavepackets with a sub-STOVs structure. A mode-locked laser source (spectral bandwidth: 20 nm, center wavelength: 1030 nm) is divided into two arms: the "object" arm and the "probe" arm. In the "object" arm (red beam path in Fig. 2), the pulse is sent into a folded two-dimensional (2D) ultrafast pulse shaper, which constitutes a reflective diffraction grating, a cylindrical lens, and a phase-only spatial light modulator (SLM) (Holoeye GAEA-2, 3840×2160 pixels with a pitch of 3.74 μm). In this configuration, the programmable SLM is positioned at the spatial-spectral plane of the input pulse for imparting a designed phase modulation. To generate the STOV wavepackets, the phase mask written into the SLM is produced using the holographic complex amplitude modulation method[33,34]. The generation process of the phase mask is demonstrated in the insets in Figure 2. The modulated beam is retroreflected and recombined by the grating to produce the designated wavepackets at the back focal plane of the Fourier lens (FL) (focal length = 1200 mm) in the setup.

In the "probe" arm (blue beam path), the pulse replica is compressed to a Fourier-transform-limited form using a grating-pair pulse compressor, and it is delivered with a controllable delay line stage. It is then sent to the CCD camera, which is also positioned in the back-focal plane of FL. At the CCD position, the "probe" and the "object" wavepackets are recombined spatiotemporally with a minor angle offset. It thus generates the interferometric pattern between these two wavepackets. By scanning the temporal delay between the pulses, the "object" wavepacket's amplitude and phase information is encoded into the resulting fringe pattern. Its three-dimensional intensity and phase profile can be later reconstructed[34].

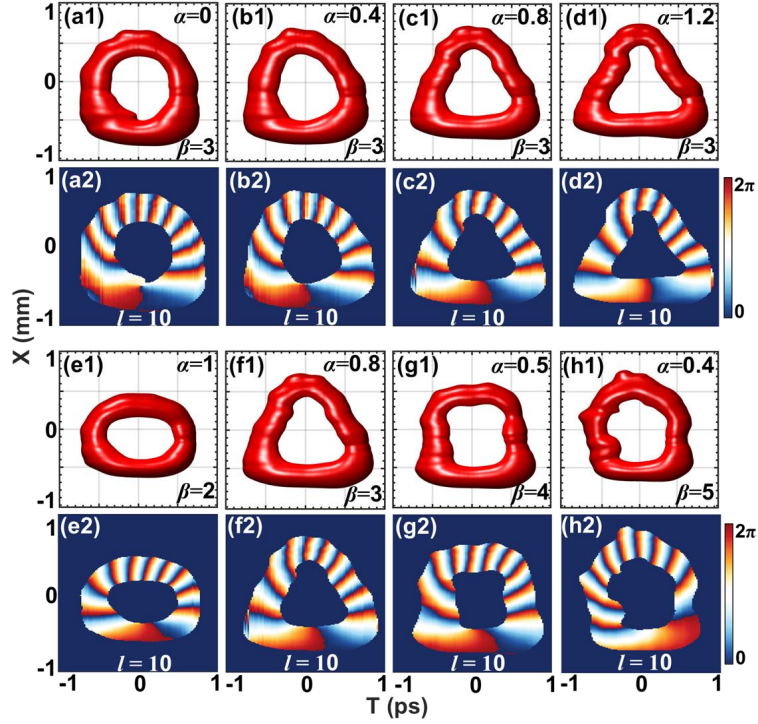

**Fig. 3. Generation of polygonal perfect STOV wavepackets with different shapes.** (a1) - (d1) Experimentally reconstructed iso-intensity profiles of the triangular perfect STOV wavepackets with different values of $\alpha$. The value of $\alpha$ corresponds to a different triangular shape for the generated spatiotemporal wavepacket. (a2) - (d2) Measured spatiotemporal phase for the triangular perfect STOV wavepackets. The carried topological charge (TC) $l$ is fixed at +10 for all cases. (e1) - (h1) Experimentally reconstructed iso-intensity profiles of the polygonal perfect STOV wavepackets. The number of sides is determined by the value of $\beta$. (e2) - (h2) Measured spatiotemporal phase for the perfect STOV wavepackets with different shapes. The carried TC $l$ is fixed at +10 for all cases.

Firstly, the effect of the nonlinear coefficient $\alpha$ was studied. The results as shown in Figs. 3(a1-d1), we experimentally reconstructed iso-intensity and phase profiles of the triangular perfect STOV wavepackets with different $\alpha$, while fixing $\beta = 3$ and $l = +10$. It can be observed that as $\alpha$ increases, the STOV wavepacket gradually changes from circular to triangular. Figure 3(a2-d2) is the corresponding measured spatiotemporal sliced phase. The desired degree of shaping can be obtained by appropriately setting the parameter $\alpha$. Furthermore, the number of sides equals the value of $\beta$ and is not related to topological charges. Therefore, by setting the appropriate parameters $(\alpha, \beta)$, we generate elliptic (1, 2), triangular (0.8,3), square (0.5,4), and pentagonal (0.4,5) STOV wavepackets. The results are shown in Figs. 3(e1-h1). These perfect STOV wavepackets have an obvious polygonal structure. Figure 3(e2-h2) is the corresponding measured spatiotemporal sliced phase.

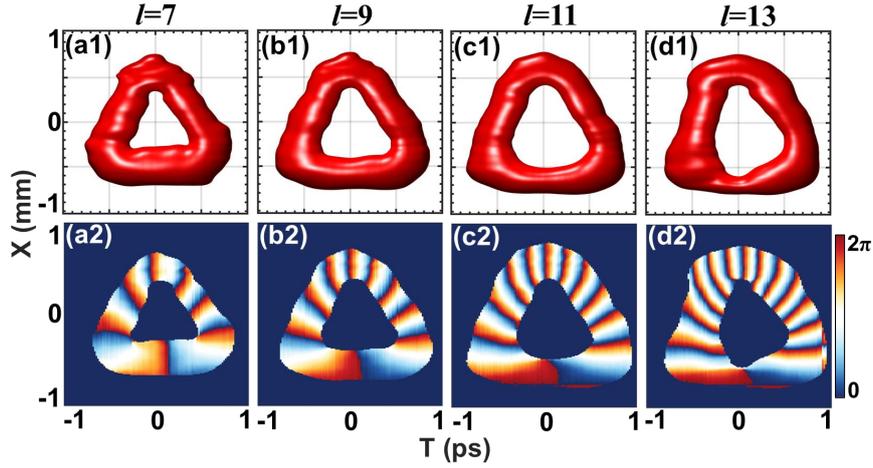

**Fig. 4. Generation of triangular perfect STOV wavepackets with different TCs.** (a1) - (d1) Experimentally reconstructed iso-intensity profiles of the triangular STOV wavepackets with different TC $l$ varying from +7 to +13. (a2) - (d2) Measured spatiotemporal phase.

Notably, the size of the perfect STOV is independent of the topological charge (TC). The polygonal perfect STOV still retains this feature, as shown in Fig. 4. As the value of TC changes from $l = +7$ to $l = +13$, the perfect STOV with a triangular shape always maintains its shape in the spatiotemporal domain.

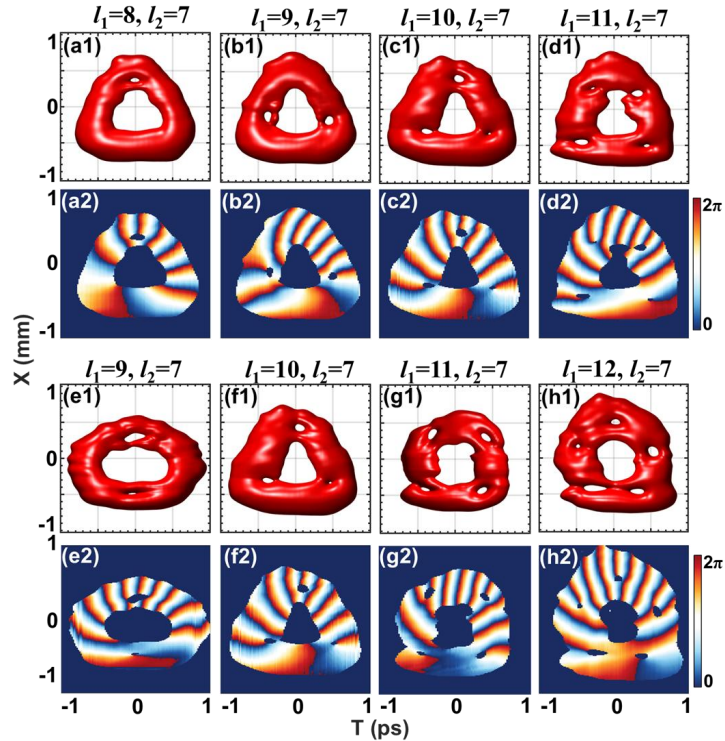

**Fig. 5. Generation of polygonal STOV wavepackets with different numbers of sub-STOVs.** (a1) - (d1) Experimentally reconstructed 3D iso-intensity profiles of the triangular STOV wavepackets with different numbers of sub-STOVs. (a2) - (d2) Measured spatiotemporal phase. (e1) - (h1) Experimentally reconstructed 3D iso-intensity profiles of the STOV wavepackets with elliptic, triangular, square, and pentagonal shapes. (e2) - (h2) Measured spatiotemporal phase.

Combined with the wavepacket shaping technique proposed above, a multi-degree-of-freedom tunable polygonal STOV wavepacket with a prescribed vortex structure can be generated by combining two polygonal STOV wavepackets centered at the same spatiotemporal position. Let $l_1$ and $l_2$ be the topological charge values of the component perfect STOV located outside and inside the STOV wavepacket with multiple sub-STOVs, respectively. We set $l_2$ unchanged and only change the value of $l_1$. The combination of the topological charges ($l_1$, $l_2$) is selected to (8,7), (9,7), (10,7), and (11,7). Figure 5(a1-d1) shows the reconstructed 3D iso-intensity profiles of the triangular STOV wavepackets with different numbers of sub-STOVs. Figure 5(a2-d2) is the corresponding measured spatiotemporal sliced phase. From the corresponding phase distribution, it can be seen that the number of sub-STOVs in these wavepackets is 1, 2, 3, 4, respectively. For each sub-STOVs, the TC is +1. The number of optical vortices satisfies the formula $N = |l_2 - l_1|$. It shows the capability of controlling the number of optical vortices in the STOV wavepacket by modifying the topological charges of the two perfect STOV wavepackets.

In addition, we can change the shape of the two component perfect STOV wavepacket to achieve the control over the shape of the combined STOV wavepacket. As shown in Figs. 5(e1-h1), we choose the parameter space $(\alpha, \beta, l_1, l_2)$ of the wavepacket as (1,2,9,7), (0.8,3,10,7), (0.5,4,11,7) and (0.4,5,12,7). The generated STOV wavepackets then have different shapes, and different numbers of sub-STOVs are generated. The shapes of these STOV wavepackets are elliptical, triangular, square, and pentagonal.

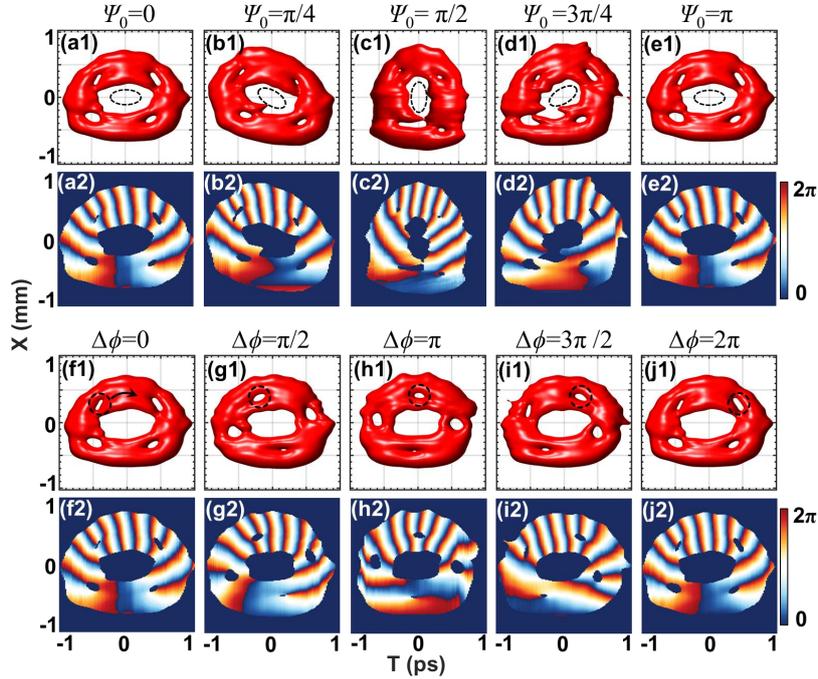

Fig. 6. Control over the elliptic STOV wavepacket rotation and the position of the sub-STOVs. (a1) - (e1) The rotation of the elliptic STOV wavepacket is achieved by varying $\Psi_0$. (a2) - (e2) Measured spatiotemporal phase. The results show the sub-STOVs are rotating together with the STOV envelope. Each sub-STOVs has a TC of +1. (f1) - (j1) Shifting the position of the sub-STOVs within the wavepacket envelope. This is achieved by changing $\Delta\phi$. (f2) - (j2) Measured spatiotemporal phase.

Compared to the circular STOV wavepacket[36], the multi-structure STOV wavepacket also possesses an additional modulation dimension, i.e., the orientation modulation. The orientation of the entire STOV wavepacket with sub-STOVs can be freely modulated by changing orientation factor $\Psi_0$ in the expression of Eqn. (1). We take the elliptical STOV wavepacket with sub-STOVs with $l_1 =+ 11$ and $l_2 =+ 7$ as an example. Figures 6(a1-e1) shows the experimentally reconstructed elliptic STOV wavepacket with orientation factor $\Psi_0$ increasing from 0 to $\pi$ by an interval of $\pi/4$. Figures 6(a2-e2) is corresponding measured spatiotemporal sliced phase. For an elliptic STOV wavepacket, the orientation repeats when the orientation factor $\Psi_0$ falls within the intervals (0, $\pi$] or ($\pi$, 2$\pi$]. In addition, the dynamic control of optical sub-STOVs can be achieved by adjusting the phase difference $\Delta\Phi = \Phi_2 - \Phi_1$ of the two components STOV. Figures 6(f1-j1) demonstrates the experimentally reconstructed elliptic STOV wavepacket with phase difference $\Delta\Phi$ increasing from 0 to $2\pi$ by an interval of $\pi/2$. The optical vortices marked with a black circle rotates clockwise by an angle of $\Delta\Phi/|l_1 - l_2|$. Figures 6(f2-j2) is the corresponding measured spatiotemporal sliced phase. Thus, by choosing an appropriate phase difference, the position of each optical sub-STOVs can be precisely controlled within the STOV wavepacket envelope. This capability enables efficient trapping and acceleration of microparticles, making it valuable for optical tweezers and related applications.

## 4. Conclusions

In summary, we theoretically propose and experimentally demonstrate a method that generates polygonal perfect STOV wavepackets with a single STOV, as well as polygonal STOV wavepackets with multiple sub-STOVs. Therein, the polygonal STOV wavepacket with multiple sub-STOVs is generated by combining two polygonal perfect spacetime vortex wave packets. Therefore, the structure and number of sub-STOVs in the wavepacket are determined by the nonlinear coefficients and TC of the two perfect STOV wavepackets respectively. Furthermore, by adjusting the orientation factor and phase difference, a variety of motions of the polygonal STOV wavepacket with a prescribed sub-STOVs structure can be realized. Such a wavepacket can bring new opportunity in new research and applications such as optical communication, multiple-target particle manipulation and studying complicated quantum systems.

**Acknowledgement.** We acknowledge financial support from National Key Research and Development Program of China [Grant No. 2025YFF0515100 (Q.Z)], National Natural Science Foundation of China (NSFC) [Grant Nos. 12434012 (Q.Z.), 62535013 (Q.Z.), and 12474336 (Q.C.)], the Shanghai Science and Technology Committee [Grant Nos. 24JD1402600 (Q.Z.) and 24QA2705800 (Q.C.)], National Research Foundation of Korea (NRF) funded by the Korea government (MSIT) [Grant No. 2022R1A2C1091890], and Global - Learning & Academic research institution for Master's·PhD students, and Postdocs (LAMP) Program of the National Research Foundation of Korea(NRF) grant funded by the Ministry of Education [No. RS-2023-00301938]. Q.Z. also acknowledges support by the Key Project of Westlake Institute for Optoelectronics [Grant No. 2023GD007].
**Disclosures.** The authors declare no conflicts of interest.
**Data availability.** Data underlying the results presented in this paper are not publicly available at this time but may be obtained from the authors upon reasonable request.